\documentclass[showpacs,preprintnumbers,amsmath,amssymb]{revtex4}

\usepackage{graphicx}
\usepackage{dcolumn}
\usepackage{bm}
\usepackage{epsf}

\newcommand{\beq}{\begin{equation}}
\newcommand{\eeq}{\end{equation}}
\newcommand\beqa{\begin{eqnarray}}
\newcommand\eeqa{\end{eqnarray}}
\newcommand\bea{\begin{array}}
\newcommand\eea{\end{array}}
\newcommand\ba{\begin{array}}
\newcommand\ea{\end{array}}
\newcommand{\nn}{\nonumber}

\newcommand{\neqa}{\nonumber\end{eqnarray}}
\newcommand{\la}{\label}
\newcommand{\vrho}{\varrho}
\newcommand{\vro}{\varrho}

\newcommand{\noi}{\noindent}
\newcommand{\eq}[1]{eq.(\ref{#1})}
\newcommand{\eqs}[2]{eqs.(\ref{#1},\ref{#2})}
\newcommand{\Eq}[1]{Eq.(\ref{#1})}

\newcommand{\ur}[1]{(\ref{#1})}
\newcommand{\urs}[2]{(\ref{#1},\ref{#2})}

\newcommand{\Tr}{{\rm Tr}}

\newcommand{\half}{\frac{1}{2}}

\renewcommand{\d}{\partial}
\renewcommand{\O}{{\cal O}}

\newcommand{\<}{{\langle}}
\renewcommand{\>}{{\rangle}}

\newcommand{\re}{\relax{\rm I\kern-.18em R}}

\def\su2{{SU(2)}}

\def\tr{{\rm tr}}

\begin{document}

\title{SU(N) caloron measure and its relation to instantons}

\author{Dmitri Diakonov$^{a,b}$}
\author{Nikolay Gromov$^c$}
\vskip 0.3true cm

\affiliation{
$^a$NORDITA,  Blegdamsvej 17, DK-2100 Copenhagen,
Denmark\\
$^b$St. Petersburg Nuclear Physics Institute, Gatchina, 188 300, St. Petersburg, Russia\\
$^c$St. Petersburg State University, Faculty of Physics, Peterhof, 198 904, St. Petersburg, Russia.
}

\date{March 12, 2005}

\begin{abstract}
Calorons of the $SU(N)$ gauge group with non-trivial holonomy, {\it
i.e.} periodic instantons with arbitrary eigenvalues of the Polyakov
line at spatial infinity, can be viewed as composed of N
Bogomolnyi--Prasad--Sommerfeld (BPS) monopoles or dyons. Using the
metric of the caloron moduli space found previously we compute
the integration measure over caloron collective coordinates in terms
of the constituent monopole positions and their $U(1)$ phases. In
the limit of small separations between dyons and/or trivial
holonomy, calorons reduce locally to the standard instantons whose
traditional collective coordinates are the instanton center, size
and orientation in the color space. We show that in this limit the
instanton collective coordinates can be explicitly written through
dyons positions and phases, and that the $N$-dyon measure coincides
exactly with the standard instanton one.
\end{abstract}

\maketitle

\section{Introduction}

Belavin--Polyakov--Schwartz--Tyupkin (BPST) instantons~\cite{BPST,Pol} are known
to play an important role in Quantum Chromodynamics, see Refs.~\cite{D96,D02} for reviews.
The instanton liquid model~\cite{ILM} is especially helpful in providing a microscopic
mechanism of the spontaneous chiral symmetry breaking, as due to the delocalization
of the would-be zero fermion modes in the instanton ensemble~\cite{DP-SCSB}.

At the same time, instantons do not lead to confinement, at least in the naive dilute limit.
In the pure glue version of QCD, there are two well-known criteria of confinement: the area
behavior of large Wilson loops, and the zero average of the Polyakov lines~\cite{Pol78}, with
its subsequent restoration to the center-of-group values at temperatures above the
deconfinement transition. To be more precise, one {\em can} formally obtain the area law for large
Wilson loops from averaging over the instanton ensemble, provided instanton size distribution
drops as $d\rho/\rho^3$ for large-size instantons~\cite{DP95}. However, such distribution
implies that large-size instantons overlap, which makes meaningless the description of
the vacuum fluctuations in terms of the instanton collective coordinates: one has to use
other degrees of freedom.

To study the temperature dependence of the average Polyakov line, one needs first to generalize the
zero-temperature BPST instantons to the periodic Harrington--Shepard instantons~\cite{HS}.
The quantum weight of periodic instantons has been found by Gross, Pisarski and Yaffe~\cite{GPY},
and the instanton ensemble at any temperatures has been built in Ref.~\cite{DMir}, using
the variational principle of Petrov and one of the authors~\cite{ILM}. Averaging the Polyakov
line over this ensemble, it was found that it oscillates near zero at small temperatures
and rapidly approaches the center-of-group value at $T\geq\Lambda$~\cite{DP-Pol-line}. However,
it is neither exactly zero at small $T$, nor is there a sharp phase transition. This is the kind
of behavior expected from the approximate treatment of large-size instantons. Again, one
concludes that in order to observe mathematically the confinement-deconfinement phase
transition, one needs to use the degrees of freedom appropriate for overlapping instantons.
It was conjectured in Ref.~\cite{Zhitn} that the adequate description should be in terms
of $N$ monopoles constituting an $SU(N)$ instanton. In a more simple $2d$ $CP^{N-1}$ model
also possessing instantons, the appropriate degrees of freedom known as ``instanton quarks"
or ``zindons" have long been available -- see Ref.~\cite{DMaul} for references and for a detailed
study of the $CP^{N-1}$ instanton ensemble in terms of their constituents.

For the $4d$ Yang--Mills theory, a somewhat similar construction of instantons through
their ``constituents'' became available more recently, owing to Kraan and van Baal~\cite{KvB}
and Lee and Lu~\cite{LL}, first for the $SU(2)$ gauge group and later for the general
$SU(N)$~\cite{KvBSUN}. These authors have found explicitly an exact self-dual solution
of the Yang--Mills equation of motion at any temperature with a unity topological charge
and with arbitrary eigenvalues of the Polyakov line (or holonomy) at spatial infinity.
We shall call this general solution the KvBLL caloron. The periodic Harrington--Shepard
instanton is a limiting case of the KvBLL caloron at trivial holonomy corresponding to
the Polyakov line assuming center-of-group values. A caloron with the double topological
charge has been constructed in Ref.~\cite{BvB}.

The fascinating feature of the $SU(N)$ calorons is that they can be viewed as composed of N
Bogomolnyi--Prasad--Sommerfeld monopoles~\cite{BPS} or, more precisely, dyons since they carry
both magnetic and electric charges; the composite calorons are electrically and magnetically
neutral. Apart from $N-1$ eigenvalues of the Polyakov line at spatial infinity, the
$SU(N)$ KvBLL caloron is characterized by $4N$ collective coordinates forming its moduli
space. A natural choice of the collective coordinates is to use $3N$ positions
of the dyons' centers in space, and $N$ dyons' $U(1)$ phases, $3N+N=4N$. If all $N$ dyons are
spatially far apart, the action density of the KvBLL caloron consists of $N$ time-independent
$3d$ lumps whose profile is the well-known energy density of individual BPS dyons.
In the opposite limit when all dyons are within the spatial range $\leq 1/T$ from each other,
the KvBLL caloron becomes a single $4d$ lump whose profile is close to the usual periodic
instanton. As the temperature goes to zero with dyons separation fixed, the caloron
action density tends to that of the standard BPST instanton. Contrary to the standard instanton,
however, the holonomy (or the Polyakov line at infinity) remains non-trivial.

The average eigenvalues of the Polyakov line are determined by the dynamics of the ensemble
of calorons with non-trivial holonomy. For example, in the ${\cal N}=1$ supersymmetric
version of the Yang--Mills theory the dyon-induced superpotential can be computed
exactly~\cite{DHKM}, and its minimum corresponds to the Polyakov line's eigenvalues
\beq
L\equiv{\rm P}\,\exp\left(\int_0^{1/T}\!dt\,A_4\right)_{|\vec  x|\to\infty}
={\rm diag}\,\left(e^{i\pi\frac{1}{N}},e^{i\pi\frac{3}{N}},...,e^{i\pi\frac{2N-1}{N}}\right),
\la{vevN}\eeq
such that $\Tr\,L=0$ as it should be in the confining phase. Moreover, the known exact
v.e.v. of the gluino condensate corresponds to this particular holonomy, whereas the
trivial-holonomy instantons lead to a wrong value~\cite{DHKM,DP-SUSY}. This result
is the more surprising that at $T\to 0$ the local difference between gauge fields with
trivial and non-trivial holonomy vanishes, implying that long-range fields are critical,
at least in the supersymmetric gluinodynamics~\cite{DP-SUSY}.

In the non-supersymmetric Yang--Mills theory, the question what average holonomy
is dynamically preferred, is open. From lattice simulations we know that in the confining
phase $<\!\Tr\,L\!>=0$ but we do not know what dynamics leads to it. Revealing
it would be tantamount to understanding the mechanism of confinement. A step
in that direction has been taken in Ref.~\cite{DGPS} where the quantum weight of the
KvBLL caloron has been computed exactly, as function of the holonomy, dyon separation
and temperature for the $SU(2)$ group. Based on this calculation, an argument has been
presented that at $T<T_c=1.125\,\Lambda$ a trivial holonomy becomes dynamically unfavorable
from the free energy minimization viewpoint. Below $T_c$ dyons repulse each other, and
calorons presumably ``ionize'' into separate dyons. However, to find out their fate and
whether the system prefers the ``confining'' holonomy \ur{vevN}, one has to study the dynamics
of {\em many} dyons. To that end one has first of all to find the statistical or quantum weight
of a dyon configuration as given by the combination of the collective coordinate measure and
the small oscillation quantum determinant.

This paper is devoted to the study of the measure of a single $SU(N)$ KvBLL caloron,
written in terms of the dyon coordinates and their $U(1)$ phases. The metric tensor
of the moduli space has been first conjectured by Lee, Weinberg and Yi~\cite{LWY,LY}
and then derived by Kraan~\cite{Kraan} using the explicit
Atiyah--Drinfeld--Hitchin--Manin--Nahm (ADHMN) construction~\cite{ADHM,Nahm80}
for the $SU(N)$ caloron~\cite{KvBSUN}. We have independently reproduced the same result
for the moduli space metric, however we do not present the derivation here as
it is lengthy but is not qualitatively different from that by Kraan. Instead, we
compute the determinant of the metric tensor, which defines the
integration measure over the dyons' collective coordinates for the
general $SU(N)$ caloron, and compare it with the long-known
instanton measure~\cite{Bernard} written in terms of the instanton
position, size and group orientation. We demonstrate that the
$SU(N)$ instanton measure written in these terms coincides exactly
with the one written in terms of the coordinates and phases of the
instanton constituents. This result is not altogether trivial, as in
the first case the measure arises from the volume of the
$SU(N)/SU(N-2)$ coset whereas in the second case it follows from the
$3d$ geometry. We also find the direct relation between the
instanton group orientation and the dyons positions and $U(1)$
phases. We believe that it may be an important step in combining the
success of the small-size instantons in physics related to the
spontaneous chiral symmetry breaking, with the description of
large-size instantons in terms of their dyon constituents, which is
presumably necessary for the confinement physics. Since this paper
concentrates mainly on the mathematical questions, we do not discuss
here the very interesting recent studies of the KvBLL calorons on
the lattice~\cite{cal-lat}.

\section{Notations}

To help navigate and read the paper, we first introduce some notations used throughout.
Basically we use the same notations as in Ref.~\cite{KvBSUN}.
In what follows we shall measure all quantities in the temperature units and put $T=1$.
The temperature factors can be restored in the final results from dimensions.

Let the Polyakov line at spatial infinity have the following eigenvalues
\beq
L={\rm P}\,\exp\left(\int_0^{1/T}\!dt\,A_4\right)_{|\vec  x|\to\infty}
=V\,{\rm diag}\left(e^{2\pi i \mu_1},\,e^{2\pi i \mu_2}\ldots e^{2\pi i \mu_N}\right)\,V^{-1},
\qquad \sum_{m=1}^N\mu_m=0.
\la{Pol1}\eeq
We use anti-hermitian gauge fields $A_\mu=it^aA_\mu^a=\frac{i}{2}\lambda^aA_\mu^a$,
$[t^at^b]=if^{abc}t^c,\,\tr(t^at^b)=\half\delta^{ab}$. The eigenvalues $\mu_m$ are
uniquely defined by the condition $\sum_{m=1}^N\mu_m=0$. If all eigenvalues
are equal up to the integer, implying $\mu_m=k/N-1,\;m\leq k$ and $\mu_m=k/N,\;m > k$
where $k=0,1,...(N-1)$, the Polyakov line belongs to the
$SU(N)$ group center, and the holonomy is then said to be ``trivial''.
By making a global gauge rotation one can always order the Polyakov line eigenvalues such that
\beq
\mu_1\leq \mu_2\leq \ldots \leq \mu_N\leq \mu_{N+1}\equiv \mu_1+1,
\la{mus}\eeq
which we shall assume done. The eigenvalues of $A_4$ in the adjoint representation,
$A_4^{ab}=if^{abc}A_4^c$, are $\pm(\mu_m-\mu_n)$ and $N-1$ zero eigenvalues.
For the trivial holonomy all adjoint eigenvalues are integers. The difference
of the neighbor eigenvalues in the fundamental representation $\nu_m\equiv\mu_{m+1}-\mu_m$
determines the spatial core size $1/\nu_m$ of the $m^{\rm th}$ monopole whose 3-coordinates
will be denoted as $\vec y_m$, and the spatial separation between neighbor monopoles
in color space will be denoted by
\beq
\vec\vrho_m\equiv\vec y_m-\vec y_{m-1}
=\vrho_m\,(\sin\theta_m\cos\phi_m,\,\sin\theta_m\sin\phi_m,\,\cos\theta_m),
\qquad\vrho_m\equiv |\vec\vrho_m|.
\la{vrho}\eeq
With each 3-vector $\vec\vrho_m$ we shall associate a 2-component spinor
$\zeta^{\dagger\,\alpha}_m$ built according to the Euler parametrization
\beq
\zeta^{\dagger\,\alpha}_m=\sqrt{\frac{\vrho_m}{\pi}}{\left[\exp\left(-i\phi_m\frac{\tau_3}{2}\right)\,
\exp\left(-i\theta_m\frac{\tau_2}{2}\right)\,\exp\left(-i\psi_m\frac{\tau_3}{2}\right)\right]_2^\alpha}
=\sqrt{\frac{\vrho_m}{\pi}}\left[\bea{c}-\sin\frac{\theta_m}{2}
\exp\left(i\frac{\psi_m-\phi_m}{2}\right)\\
\cos\frac{\theta_m}{2}\exp\left(i\frac{\psi_m+\phi_m}{2}\right)\eea\right]^\alpha.
\la{param}\eeq
This spinor, together with its Hermitian conjugate $\zeta^m_\alpha$, forms a $2\times 2$ matrix
for any $m=1...N$:
\beq
\zeta^{\dagger\,\alpha}_m\zeta^m_\beta=\frac{1}{2\pi}
\left(1_2\vrho_m-\vec\tau\cdot\vec\vrho_m\right)^\alpha_\beta\,.
\la{eq2}\eeq
These spinors are used in the construction of the caloron field. The Euler angle $\psi_m$ is fictitious
in parameterizing the $3d$ vector $\vec\vrho_m$ but enters explicitly the gauge field of
the caloron and belongs to its moduli space, together with $\vec\vrho_m$. In fact $\psi_m$
has the meaning of the $U(1)$ phase of the $m^{\rm th}$ dyon. We shall also use the following
notation for the variation
\beq
\frac{i\pi}{\vrho_m}\tr(\zeta^\dag_m\delta\zeta^m-\delta\zeta^\dag_m \zeta^m)=
\delta\psi_m+\cos\theta_m \delta\phi_m \equiv \delta\Sigma_m.
\la{defSigma}\eeq

For trivial holonomy, the KvBLL caloron reduces to the Harrington--Shepard periodic instanton
at non-zero temperatures and to the ordinary Belavin--Polyakov--Schwartz--Tyupkin instanton
at zero temperature. Instantons are usually characterized by the scale parameter (the ``size''
of the instanton) $\rho$. It is directly related to the dyons positions in space, actually
to the perimeter of the polygon formed by dyons,
\beq
\rho=\sqrt{\frac{1}{2\pi T}\sum_{m=1}^N \vrho_m}\,,\qquad \sum_{m=1}^N \vec\vrho_m=0.
\la{size}\eeq

In these notations the KvBLL caloron gauge field can be written as
the following $N\times N$ matrix~\cite{KvBSUN}:
\beq
A^{mn}_\mu=\frac{1}{2}\phi_{mk}^{1/2}\zeta^k_\alpha
\bar\eta^a_{\mu\nu}(\tau^a)^\alpha_\beta\zeta^{\dag\,\beta}_l\d_\nu
f_{kl}\phi_{ln}^{1/2}
+\frac{1}{2}\left(\phi^{1/2}_{mk}\d_\mu\phi_{kn}^{-1/2}-\d_\mu\phi_{mk}^{-1/2}\phi_{kn}^{1/2}\right)
\la{Amu2}\eeq
where the summation over $k,l$ is understood and where
\beq
{\phi^{-1}}_{mn}=\delta_{mn}-\zeta^m_\alpha\zeta^{\dag\,\alpha}_n
f_{mn}\,.
\la{defphi}\eeq
The $N\times N$ matrix $f_{mn}$ is in fact the ADHMN Green function $f(\mu_n,\mu_m)$
found in~\cite{CKB}. In Appendix B we derive a simple expression for this quantity used to
obtain certain limiting cases of the general \eq{Amu2}.

\section{Zero modes in the Yang-Mills theory}

Here we remind what are zero modes and how the moduli space metrics arises from the path integral.
In our notations the partition function for the pure Yang-Mills theory reads
\beqa
Z=\int DA \exp(-S[A]),\;\;\;\;\;
S[A]=-\frac{1}{2g^2}\int d^4x\;\tr F_{\mu\nu}F_{\mu\nu}\,,
\la{Z}\eeqa
where $A_\mu(x_4,\vec x)$ must obey the periodicity condition
$A_\mu(0,\vec x)=A_\mu(1/T,\vec x)$.

The integration measure in \eq{Z} is defined through the scalar product
\beq
\< u,u'\>=-2\int d^4x\;\tr\left(u_\mu(x)u'_\mu(x)\right)
\la{eq:metric}\eeq
by
\beq
DA=\prod_n\frac{d \alpha_n}{\sqrt{2\pi}g}
\eeq
where $A_\mu(x)=\sum_n \alpha_n u_{n\,\mu}(x)$ for the complete normalized set of
functions $u_{n\,\mu}(x)$.

We want to compute the contribution to the partition function from some set of solutions
of the classical Yang-Mills equation of motion $A_\mu(Y,x)$ parameterized by the collective
coordinates $Y_p$. It means that we have to take into account only small fluctuations
about the surface formed by this set of solutions in the configuration space.
Usually $S[A(Y)]=S^{\rm cl}$ is the same for the whole set, and is locally minimal.
The integral over fluctuations is Gaussian only in the directions orthogonal to the
surface. We have to separate Gaussian and non-Gaussian variables of integration.

The result in the quadratic order after fixing the background gauge $D_\mu^{\rm cl}a_\mu=0$ is
\beqa
&& Z_{A(Y)}=e^{-S^{\rm cl}}\int J \prod_p\frac{dY_p}{\sqrt{2\pi}g}\int Da_\mu D\chi D\bar\chi
\exp\left(-\frac{1}{2g^2}\int d^4x\; a_\mu^a W^{ab}_{\mu\nu} a^b_\nu
-\int d^4x\;\bar\chi^a (D^2)^{ab}\chi^b\right)
\la{Zcl}
\eeqa
where $a_\mu=i t^b a^b_\mu$ ($\tr(t^at^b)=\frac{1}{2}\delta^{ab}$) are small fluctuations
orthogonal to the zero modes of the operator $W_{\mu\nu}$,
\beq
W_{\mu\nu}^{ab}=-D^2[A(Y)]^{ab}\delta_{\mu\nu}-2f^{acb}F^c_{\mu\nu}[A(Y)],
\eeq
$\chi$ and $\bar\chi$ are the ghost fields from gauge fixing.
The factor $\sqrt{2\pi}g$ comes from the definition of the measure.
The Jacobian $J$ is in fact the determinant of the moduli space metric tensor, {\it i.e.}
\beq
J=\sqrt{\det g_{pq}},\qquad g_{pq}=\left\langle \delta_p A_\mu \delta_q A_\mu\right\rangle,
\la{defg}\eeq
where $\delta_p A_\mu$ is a zero-mode of $W_{\mu\nu}$, associated with
the collective coordinate $Y_p$ through
\beq
\delta_p A_\mu=\d_{Y_p}A_\mu+D_\mu\Omega_p
\la{Omega}\eeq
where $\Omega_p$ is chosen such that the background gauge condition is satisfied,
\beq
D_\mu\delta_p A_\mu=0.
\la{zmeqn}
\eeq

In the next section we present the result for the metric tensor $g_{pq}$ for the KvBLL
caloron of the $SU(N)$ gauge group.

\section{Caloron moduli space metric}

As mentioned in the Introduction, the metric of the moduli space of $N$ different BPS monopoles
of the $SU(N)$ gauge group has been first conjectured in Refs.~\cite{LWY,LY} generalizing the
previous work~\cite{GM}, and then confirmed by an explicit calculation in Ref.~\cite{Kraan}.
In these papers, the metric tensor was expressed in terms of the monopole
-- electric charge interaction potential $w_i({\bf \vrho})$ satisfying the equation
\beq
\epsilon_{ijk}\,\partial_jw_k(\vec\vrho)=\partial_i\frac{1}{\vrho}=-\frac{\vrho_i}{\vrho^3}.
\la{mon-charge-1}\eeq
One introduces also a $N\times N$ matrix
\beq
S=\left(\ba{cccccc}1&-1& & & &\\ &1&-1& & &\\ & & &...& &\\ & & & &1&-1 \\-1 & & & & & 1\ea\right),\qquad
S^T=\left(\ba{cccccc}1& & & & &-1 \\-1&1& & & & \\ & & &...& &\\ & & &-1&1 & \\ & & & &-1 & 1\ea\right),
\la{SST}\eeq
such that the separation between consecutive dyons is $\vec \rho_m=\vec y_m-\vec y_{m-1}=(S^T\vec y)_m$.
In terms of the dyon interaction potential $\vec w(\vec\vrho)$ the metric found in
Refs.~\cite{LWY,LY,Kraan} is (see e.g. Eq. (78) in~\cite{Kraan}~\cite{footnote1})
\beq
ds^2=8\pi^2\left[d\vec y^{\,T}\,G\cdot d\vec y+\left(\frac{d\tau}{4\pi}
+\vec{\cal W}\cdot d\vec y\right)^{\!T}\,G^{-1}\,
\left(\frac{d\tau}{4\pi}+\vec{\cal W}\cdot d\vec y\right)\right]
\la{metric_Kraan}\eeq
where
\beqa
\la{ob1}
&&\vec{\cal W}=S\vec W S^T,\qquad
\vec W = \frac{1}{4\pi}{\rm diag}(\vec w(\vec\vrho_1),...,\vec w(\vec\vrho_N))\,,\\
\la{ob2}
&&G={\cal N}+\frac{1}{4\pi}SR^{-1}S^T,\qquad {\cal N}={\rm diag}(\nu_1,...,\nu_N)\,,\qquad
R={\rm diag}(\vrho_1,...,\vrho_N)\,,\\
\la{ob3}
&&\frac{d\tau_m}{4\pi}=\nu_m d\xi_4+\frac{1}{4\pi}(S\psi)_m\,.
\eeqa

The metric \ur{metric_Kraan} is implicit as it employs the notion of the
monopole--electric charge interaction potential $\vec w$ which by itself is ambiguous
as it does not exist without a Dirac string singularity. The combination $(\vec w(\vrho)\cdot d\vec\vrho)$
is however independent of the way one introduces the Dirac string singularity.
Choosing it along the $z$ axis and solving \eq{mon-charge-1} we find the
monopole--electric charge interaction potential
\beq
\vec w(\vec\vrho)=\frac{1}{\vrho}(-\cot\theta\,\sin\phi,\,\cot\theta\,\cos\phi,\,0)
\la{mon-charge-2}\eeq
if one parameterizes $\vec\vrho=(\sin\theta\,\cos\phi,\,\sin\theta\,\sin\phi,\,\cos\theta)$.
Hence, $\vec{\cal W}\cdot \vec{dy}$ in \eq{metric_Kraan} can be rewritten as
\beq
\vec{\cal W}\cdot \vec{dy}=S\,\vec W\cdot d\vec\vrho=\frac{1}{4\pi}\,S\,{\rm diag}
(\cos\theta_1d\phi_1,...,\cos\theta_Nd\phi_N)\,.
\la{p1}\eeq
Being combined with \eq{ob3} it gives
\beq
\frac{d\tau}{4\pi}+\vec{\cal W}\cdot d\vec y={\cal N}d\xi_4+\frac{1}{4\pi}S\,d\Sigma
\la{p2}\eeq
where $d\Sigma_m=\cos\theta_m d\phi_m+d\psi_m$ according to \eq{defSigma}. Therefore, the
second term in \eq{metric_Kraan} can be written as
\beq
\left({\cal N}d\xi_4+\frac{1}{4\pi}S\,d\Sigma\right)^T\,G^{-1}\,
\left({\cal N}d\xi_4+\frac{1}{4\pi}S\,d\Sigma\right)
\la{p3}\eeq
where, according to \eq{ob2},
\beq
G^{-1}=\left({\cal N}+\frac{1}{4\pi}SR^{-1}S^T\right)^{-1}
={\cal N}^{-1}-{\cal N}^{-1}\,\frac{1}{4\pi}SR^{-1}S^T\,{\cal N}^{-1}+...
\la{p4}\eeq
Since $d\xi_4$ is, in this context, a $N$-vector $(d\xi_4,...,d\xi_4)$, one has
for all components $(d\xi_4\,S)_m=0$, see the definition of $S$ in \eq{SST}. Hence,
the term quadratic in $d\xi_4$ in \eq{p3} is simply
\beq
{\cal N}d\xi_4\,G^{-1}\,{\cal N}d\xi_4 = d\xi_4^2
\la{p5}\eeq
where $\tr\,{\cal N}=\sum_m\nu_m=1$ has been used. Because of $(d\xi_4\,S)_m=0$, the
terms linear in $d\xi_4$ in \eq{p4} are zero. In the last term, quadratic in $d\Sigma$,
we note that
\beq
S^TG^{-1}S=S^T{\cal N}^{-1}S-\frac{1}{4\pi}S^T{\cal N}^{-1}SR^{-1}{\cal N}^{-1}S+...
=4\pi R(4\pi R+S^T{\cal N}^{-1}S)^{-1}S^T{\cal N}^{-1}S\,.
\la{p6}\eeq
We introduce $N\times N$ matrices $K,L,M$:
\beqa\la{K1}
&&K=\frac{1}{\pi}R+\frac{1}{4\pi^2}S^T{\cal N}^{-1}S,\\
\la{L1}
&&L=\pi R-RK^{-1}R=RK^{-1}(\pi K -R)=RK^{-1}\frac{1}{4\pi}S^T{\cal N}^{-1}S,\\
\la{M1}
&&M=4\pi^2 G.
\eeqa
With the help of these matrices, the chain of \eq{p6} can be continued as
\beq
S^TG^{-1}S=4\pi R\frac{1}{4\pi^2}K^{-1}\,4\pi(\pi K-R)=4L.
\la{p7}\eeq
Thus, the last term in \eq{p3} is
\beq
\frac{1}{4\pi^2}d\Sigma\,L\,d\Sigma.
\la{p8}\eeq
Combining all terms from \eq{metric_Kraan} we obtain finally a simple and explicit
expression for the moduli space metric:
\beq
ds^2= 8\pi^2d\xi_4^2+2M_{mn} d\vec y_m\, d\vec y_n+2L_{mn}\,d\Sigma_md\Sigma_n,
\la{ds1}\eeq
where $d\Sigma_m$ is given by \eq{defSigma}.
As a matter of fact, we have independently derived the moduli space metric in precisely
this form, using the ADHMN construction for the $SU(N)$ caloron. However, since
the derivation is lengthy but not qualitatively different from that of Kraan~\cite{Kraan}
we do not present it here. Explicitly, the $K,L,M$ matrices involved in \eq{ds1} are
\beqa &K_{mn}=\left(\frac{\vrho_n}{\pi}
+\frac{1}{4\pi^2\nu_n}+\frac{1}{4\pi^2\nu_{n-1}}\right)\delta_{mn}
-\frac{1}{4\pi^2\nu_m}\delta_{m+1,n}-\frac{1}{4\pi^2\nu_{n}}\delta_{m,n+1},&
\la{K}\\
\la{L}
&L_{mn}=\pi R_{mn}-(R K^{-1} R)_{mn},\;\;\;\;\;R_{mn}\equiv \delta_{mn}\vrho_n, &\\
\la{M}
&M_{mn}\equiv\left(4\pi^2\nu_n+\frac{\pi}{\vrho_n}+\frac{\pi}{\vrho_{n+1}}\right)\delta_{mn}
-\frac{\pi}{\vrho_{n}}\delta_{m+1,n}-\frac{\pi}{\vrho_m}\delta_{m,n+1}\,.&
\eeqa
Notably $K$ and $M$ are symmetric and differ only by interchanging
$4\pi^2\nu_m$ and $\rho_m/\pi$: this will be used in computing the determinants.

As an example, we give the matrix $M$ for the $SU(4)$ gauge group:
\beq
M^{(4)}=\pi\left[\bea{cccc}
4\pi \nu_1+\frac{1}{\vrho_1}+\frac{1}{\vrho_2}&-\frac{1}{\vrho_2}&0&-\frac{1}{\vrho_1}\\
&&&\\
-\frac{1}{\vrho_2}&4\pi \nu_2+\frac{1}{\vrho_2}+\frac{1}{\vrho_3}&-\frac{1}{\vrho_3}&0\\
&&&\\
0&-\frac{1}{\vrho_3}&4\pi \nu_3+\frac{1}{\vrho_3}+\frac{1}{\vrho_4}&-\frac{1}{\vrho_4}\\
&&&\\
-\frac{1}{\vrho_1}&0&-\frac{1}{\vrho_4}&4\pi \nu_4+\frac{1}{\vrho_4}+\frac{1}{\vrho_1}\eea\right].
\la{M4}\eeq

The $SU(2)$ gauge group is too ``small'' for the general formula \ur{M}. In this case the
matrix $M$ is simply
\beq
M^{(2)}=\pi\left[\bea{cc}
4\pi \nu_1+\frac{1}{\vrho_1}+\frac{1}{\vrho_2}&-\frac{1}{\vrho_1}-\frac{1}{\vrho_2}\\
&\\
-\frac{1}{\vrho_1}-\frac{1}{\vrho_2}&4\pi \nu_2+\frac{1}{\vrho_1}+\frac{1}{\vrho_2}\eea\right]
\la{M2}\eeq
where $\vrho_1=\vrho_2=|\vec y_1-\vec y_2|$ and $\nu_1+\nu_2=1$.

\section{The determinant of the metric tensor}

In the previous section we have rewritten the moduli space metric in the explicit form \ur{ds1}.
However only the determinant of the metric is needed in such calculations as
the saddle point approximation, see. \eq{Zcl}. In this section we derive
a compact expression for the volume of the general $SU(N)$ moduli space and then
give examples for the specific cases of the $SU(2)$ and $SU(3)$ groups,
as well as an asymptotic formula for the general $SU(N)$ group, valid at large
separations between the dyons.

First of all, we need to check the dimension or the number of parameters of the moduli space.
These are the $3N$ coordinates of dyon centers $\vec y_i$, one overall time position $\xi_4$, and
$N-1$ relative color orientations $\psi_m$ entering the metric \ur{ds1}
from \eq{defSigma}. Therefore, the dimension of the caloron moduli space is $4N$ as it should be
for a general self-dual solution with unity topological charge. We note that the transformation
$\delta\psi_1=\delta\psi_2=\ldots =\delta\psi_N$ is a global $U(1)$ gauge rotation
leaving the gauge field unchanged. As a consequence, the matrix $L$ has one zero eigenvalue
\beq
L|1,\dots,1>=0
\eeq
which makes the size of the maximal non-degenerate minor of the metric tensor (\ref{ds1}) equal to
$4N$. The determinant of the metric tensor is
\beq
g\equiv\det g_{pq}=8\pi^2 2^{3N}{\det}^3 M\; 2^{N-1}{\det}' L
\eeq
where ${\det}'L$ is the product of all non-zero eigenvalues of $L$. The corresponding volume form is
\beq
\omega=2^{N-1}\sqrt{N}\sqrt{g}\;d\xi_4\; d^3y_1\dots d^3y_{N}\;d\alpha_1 \dots d\alpha_{N-1}\la{omega1}
\eeq
where
\beqa
\la{alphapsi}
\alpha_m&\equiv&\frac{\psi_m}{2}-\sum_{n=1}^N\frac{\psi_{n}}{2N},\;\;\;\;\;m=1,\dots,N-1,\\
\nn\alpha_N&\equiv&\sum_{n=1}^N\frac{\psi_{n}}{2N}
\eeqa
is a set of variables that parameterize the relative $U(1)$ orientations of the dyons.
Note that $\alpha_N$ corresponds to the trivial gauge transformation. The transformation matrix has the form
\beq\nn
Q_{mn}\equiv\frac{d\psi_m}{d\alpha_n}=
\frac{1}{2}\left(
\bea{cccc}
\frac{N-1}{N} & -\frac{1}{N} & \ldots & -\frac{1}{N}\\
-\frac{1}{N} & \frac{N-1}{N} & \ldots & -\frac{1}{N}\\
\vdots & \vdots  &\ddots & \vdots\\
\frac{1}{N} & \frac{1}{N} & \ldots & \frac{1}{N}
\eea
\right)^{-1}
\eeq
The factor $2^{N-1}\sqrt{N}$ in eq.(\ref{omega1})
comes from the equation
\beqa
{\det}'(Q^T L Q)&=&\lim_{\epsilon\rightarrow 0}\frac{\det(Q^T L Q+\epsilon)}{\epsilon}
=\det(QQ^T)\lim_{\epsilon\rightarrow 0}\frac{\det(L+(Q Q^T)^{-1}\epsilon)}{\epsilon}\\
&=&\det(QQ^T)(QQ^T)^{-1}_{00}{\det}'(L)=(2^{N-1}\sqrt{N})^2 {\det}'L
\eeqa
where $(QQ^T)^{-1}_{00}\equiv \langle 1,1,\dots,1|(QQ^T)^{-1}|1,1,\dots,1\rangle/N$.

It turns out that ${\det}'L$ can be expressed through $\det K$. To show this let us introduce
\beq
V\equiv 4\pi\left(\pi K-R\right)
\eeq
such that the matrix $L$ from \eq{L} can be written as
\beqa
&&L\equiv \pi R-R K^{-1}R=RK^{-1}(\pi K-R)=\frac{1}{4\pi}R K^{-1}V\,.
\la{RKV}\eeqa
Using that $<1,1,\dots,1|RK^{-1}=\pi<1,1,\dots,1|$ we have
\beq
{\det}'L=4^{1-N}\frac{\det R\;{\det}' V}{\pi^{N}\det K}
\eeq
and from a simply calculable ${\det}'V=\frac{N}{\prod_{m=1}^{N}\nu_m}$ we obtain
\beq
{\det}'L=\frac{N}{2^{2N-2}\pi^N\det K}\frac{\prod\vrho_m}{\prod \nu_m}\,.
\eeq
In its turn, $\det K$ has a simple relation to $\det M$
\beq
\frac{\det M}{\det K}=\frac{\prod 4\pi^2\nu_m}{\prod \frac{\vrho_m}{\pi}}
\eeq
which follows from the symmetry between the two matrices mentioned at the end of Section IV.B.
Thus, the final result for the element of the volume of the moduli space is
\beq
\omega=(4\pi)^{N+1}2^{N-1}N\det M\; d\xi_4\;d^3y_1\dots d^3 y_N\;d\alpha_1\dots d\alpha_{N-1}
\la{vol1}\eeq
where $M$ is given by \eq{M}. One can also rewrite it in terms of the ``center of mass'' position
$\vec\xi=\sum_m \nu_m \vec y_m$ and the separations between dyons neighboring in color space, $\rho_m$:
\beq
\omega=(4\pi)^{N+1}2^{N-1}N\det M\; d^4\xi\;d^3\rho_1\dots d^3 \rho_{N-1}\;d\alpha_1\dots d\alpha_{N-1}\,.
\la{sqrtg}\eeq

\section{Integration over dyons' $U(1)$ phases}

Since for a single-charged caloron the volume element does not depend on the $U(1)$ phases
of the dyons $\psi_m$ or, equivalently, $\alpha_m$, these phases can be integrated out.
Fortunately the integration limits in $\alpha_m$ variables are simple.
These variables parameterize a general diagonal $SU(N)$ matrix:
\beq
U(\alpha_m)={\rm diag}\{e^{i\alpha_1},e^{i\alpha_2},\dots,e^{i\alpha_{N-1}},e^{-i\sum\alpha_m}\}\,.
\eeq
It is clear that $U(\alpha_m)=U(\alpha_m')$ if and only if $\alpha'_m=\alpha_m+2\pi n_m$
where $n_m$ are integers. However $U(\alpha_m)$ and $U(\alpha_m')$ can differ by an element of the
centre of $SU(N)$ i.e. $U(\alpha_m)=U_Z U(\alpha_m')$ where $U_Z=e^{2\pi i/N}1_N$. Since $U_Z$
acts trivially in the adjoint representation we have to choose the fundamental domain of integration,
such that if $\alpha_m$ and $\alpha_m'$ are elements of this domain, the condition
$U(\alpha_m)=U_Z U(\alpha_m')$ implies that $\alpha_m=\alpha_m'$. For example, one can
choose the fundamental domain to be
\beq
0\leq\alpha_1<\frac{2\pi}{N},\qquad 0\leq\alpha_{m>1}<2\pi.
\la{alpha}\eeq
We now integrate over $\alpha_m$ in the fundamental domain specified by \ur{alpha} and obtain
\beq
\int d^{N-1}\alpha_m
=(2\pi)^{N-1}\frac{1}{N}.
\la{intsi}\eeq
Thus, the caloron measure integrated over the $U(1)$ phases (denoted by $\cal G$) is
\beq
\int_{\cal G}\omega =(4\pi)^{2N}\det M\; d^4\xi\;d^3\rho_1\dots d^3 \rho_{N-1}\,.
\la{sqrtg2}\eeq

Below we find $\det M$ in the particular cases of the $SU(2)$ and $SU(3)$ gauge groups
and in the limit of large dyon separations in a general $SU(N)$ case.

\subsection{SU(2)}

The general expression for the $N\times N$ matrix $M$ is given in \eq{M}. Computing its determinant
in the case of $N=2$ we get
\beq
\omega=2^{9}\pi^6 \frac{1+2\pi\vrho_1\nu_1\bar\nu_1}{\vrho_1} d^4\xi d^3\vrho_1 d(\psi_1-\psi_2)
\eeq
where we use the notation $\bar \nu_m=1-\nu_m$. Integrating over the $U(1)$ phase $(\psi_1-\psi_2)$
and over space rotations we get
\beq
\int\!d^3\vrho_1 d(\psi_1-\psi_2) =(4\pi)^2\int\!\vrho_1^2 d\vrho_1
=2^{4}\pi^2\int\!\vrho^2_1 d\vrho_1.
\eeq
According to \eq{alphapsi} $\psi_1-\psi_2=4\alpha_1$ and thus $\int d(\psi_1-\psi_2)=4\int_0^\pi d\alpha_1=4\pi$.
Replacing $\vrho_1$ by the commonly used instanton size variable according to \eq{size},
$\vrho_1=\pi\rho^2$, we arrive to the result already known in the $SU(2)$ case~\cite{KvB,DGPS}
\beq
\int_{{\cal G},\;{\rm rotations}}\omega
=2^{14}\pi^{10} (1+2\pi^2\rho^2\nu_1\bar\nu_1)\rho^3 d\rho\,d^4\xi.
\la{SU2}\eeq
At trivial holonomy ($\nu_1=0$) it becomes the well-known measure of the (periodic) $SU(2)$ instanton.

\subsection{SU(3)}

Computing the determinant of the $3\times 3$ matrix $M$ \ur{M} and putting it into \eq{sqrtg2}
we obtain the caloron measure for $SU(3)$:
\beqa\nn
\int_{\cal G}\omega=2^{14}\pi^{10}\left\{16\pi^2\nu_1\nu_2\nu_3
+4\pi\left[\frac{\nu_2\bar\nu_2}{\vrho_1}+\frac{\nu_3\bar\nu_3}{\vrho_2}
+\frac{\nu_1\bar\nu_1}{\vrho_3}\right]
+\left(\frac{1}{\vrho_1\vrho_2}+\frac{1}{\vrho_2\vrho_3}+\frac{1}{\vrho_3\vrho_1}\right)\right\}
d^3\vrho_1d^3\vrho_2\, d^4\xi.
\la{SU31}\eeqa

\subsection{SU(N), large separations}

In the general case, $\det M$ cannot be written in an easy form. However, for large
separations between dyons in a caloron one can derive a simple asymptotic for $\det M$,
\eq{M}. We expand it in inverse powers of $\vrho_m$. Let us write
\beq
M=4\pi^2 {\cal N}+\pi M_1,\qquad {\cal N}_{nm}\equiv\delta_{nm}\nu_m,
\eeq
where the matrix $M_1$ is composed of the inverse powers of $\vrho_m$. We have
\beq
\det M=\det(4\pi^2 {\cal N})\exp\tr\log\left(1+\frac{{\cal N}^{-1}M_1}{4\pi}\right)
\simeq (2\pi)^{2N}\left(1+\sum_m\frac{1}{4\pi\vrho_m\nu_m}
+\frac{1}{4\pi\vrho_m\nu_{m-1}}\right)\prod_n\nu_n.
\eeq
Hence from \eq{sqrtg2} we obtain the caloron measure
\beq
\int_{\cal G}\omega\simeq 2^{6N}\pi^{4N}\left[1+\sum_m\frac{1}{4\pi\vrho_m}
\left(\frac{1}{\nu_{m-1}}+\frac{1}{\nu_m}\right)\right]\prod_n\nu_n\;
d^3\vrho_1\dots d^3\vrho_{N-1}\; d^4\xi.
\la{SUN}\eeq
We remind the reader that periodicity in the indices is assumed; for example $\nu_N\equiv\mu_{N+1}-\mu_N
=\mu_1+1-\mu_N$.

\Eq{SUN} can be interpreted as Coulomb repulsion of dyons inside a caloron. However, not all
dyons interact with each other but only those that are ``neighbors'' in the color space. In $SU(3)$
all dyons are neighbors in this sense, while the $SU(2)$ group is too small to see the effect.

\section{Relation to the instanton measure in the trivial holonomy limit}

In this limit, the KvBLL caloron becomes the Harrington-Shepard periodic instanton with
the standard BPST instanton moduli space. It is basically an $SU(2)$ configuration
embedded into the $SU(N)$ group. The $4N$-parameter moduli space is usually described
as 4 `center-of-mass' coordinates $\xi_\mu$, one `size' parameter $\rho$, and
$4N-5$ `gauge orientation' collective coordinates determining the embedding. This
has been the traditional parametrization of instantons for 25 years, starting from
the work by Bernard~\cite{Bernard} who computed the instanton measure and its volume
for a general $SU(N)$ group.

At first glance, there is little in common between this moduli space and that of
the non-trivial caloron, given in the previous sections in terms of the constituent
dyons' $3d$ positions and $U(1)$ phases. Our goal is to demonstrate that the measures of
the two moduli spaces in fact coincide exactly, including the non-trivial normalization.

We shall do it in two steps. In this section we show that the volume of the dyon moduli
space coincides with that found by Bernard in terms of the $SU(2)$ embedding. In the next
section we give an explicit construction of the instanton $SU(N)$ gauge orientation
matrix (determining the $SU(2)$ embedding) in terms of the $3d$ positions and $U(1)$
phases of the constituent dyons.

The trivial holonomy limit corresponds to taking all Polyakov eigenvalues equal
$\mu_m=k/N-1,\;m\leq k$ and $\mu_m=k/N,\;m> k,$ where $k=0,1,...,(N-1)$,
meaning all their differences $\nu_m=\mu_{m+1}-\mu_m$ are zero except one which is unity,
see section II. First of all we note that in this limit one gets
\beq
\det M=4\pi^{N+1}\frac{s}{\prod_m\vrho_m},\qquad s=\sum_{m=1}^N\vrho_m,
\la{M_triv}\eeq
where $s$ is the perimeter of the polygon formed by the dyons. It is directly related
to the instanton size $\rho$ by \eq{size}: $\rho=\sqrt{s/2\pi}$.
To find the volume of the dyons moduli space and relate it to the standard instanton one,
we have to integrate \eq{sqrtg2} over dyons' $3d$ positions with the perimeter $s$ fixed.
More concretely, we have to evaluate
\beq
\mu_N(s)\equiv\int \prod_{i=1}^{N}d^3\vrho_m\det M\;\delta\!\left(\sum_{i=1}^N\vrho_m-s\right)
\delta^3\!\left(\sum_{i=1}^N\vec\vrho_m\right)
=\int \prod_{i=1}^{N}d^3\vrho_m\frac{4\pi^{N+1} s}{\prod_m\vrho_m}\;
\delta\!\left(\sum_{i=1}^N\vrho_m-s\right)\;\delta^3\!\left(\sum_{i=1}^N\vec\vrho_m\right).
\la{defmu1}\eeq
Leaving unintegrated the center of mass 4-coordinate $\xi_\mu$ and the instanton size $\rho$,
the moduli space volume is, from \eq{sqrtg2},
\beq
\int_{\cal G}\omega=\int(4\pi)^{2N}\mu_N(s)\, ds\, d^4\xi
\la{eq96}\eeq
The integral \ur{defmu1} is computed in Appendix A with the result
\beq
\mu_N(s)=\frac{2^3\pi^{2N}s^{2N-3}}{(N-1)!(N-2)!}\;.
\la{res12}
\eeq
Consequently
\beq
\int_{\cal G}\omega=\int\frac{2^{6N+2}\pi^{6N-2}}{(N-1)!(N-2)!}\,\rho^{4N-5}\,d\rho\, d^4\xi
\la{sres}
\eeq
coinciding exactly with Bernard's result~\cite{Bernard}. It is interesting to note that
it was obtained there in a completely different way -- by computing the group volume
for the embedding of $SU(2)$ into $SU(N)$. There seems to be nothing near it in the present
derivation.

\section{Limiting cases of the caloron gauge field}

In this section we give the trivial holonomy limit of the
KvBLL gauge field. It is the Harrington-Shepard $SU(2)$ instanton imbedded
into $SU(N)$. The way it is embedded depends on the constituent dyons color
orientations and their relative positions. As a byproduct, we give the gauge field
of the KvBLL caloron with the exponential precision (i.e. dropping terms of the
order of $\O(e^{-2\pi r_m\nu_m})$, where $r_m$ is a distance to the $m^{\rm th}$ dyon).

\subsection{Far from the cores}

With the exponential precision, the matrix $F_{mn}$ (see \eq{F}) is diagonal and so is the matrix $f_{mn}$:
\beq
f_{mn}= 2\pi \delta_{mn}(r_m+r_{m-1}+\vrho_m)^{-1}+\O(e^{-2\pi r_m\nu_m})\,.
\eeq
From \eq{eq2} one has
\beq
f_{mn}\zeta^\dag_m\zeta^m=\frac{\vrho_m}{\pi} f_{mn},\;\;\;\;\;
f_{mn}\zeta^m\bar\eta_{\mu\nu}\zeta_m^\dag=-\frac{\vrho^a_m}{\pi}f_{mn}\bar\eta_{\mu\nu}^a
\eeq
and from \eq{defphi}
\beq
\phi_{mn}\simeq\delta_{mn}\frac{r_m+r_{m-1}+\vrho_m}{r_m+r_{m-1}-\vrho_m}\,.
\eeq
The last term in \eq{Amu2} is zero, and with the exponential precision we can write the gauge field
\beq
A^{mn}_\mu\simeq\frac{1}{2}(\phi\d_\nu\lambda\bar\eta_{\mu\nu}f\lambda^\dag)_{mn}
\simeq-\frac{\vrho_m^a}{2\pi}\bar\eta^a_{\mu\nu}\phi_{mk}\d_\nu f_{kn}
\simeq\bar\eta^a_{\mu\nu}\frac{\vrho_m^a}{2\vrho_m}\phi_{mk}\d_\nu \phi^{-1}_{kn}\,.
\eeq
This expression is similar to the one found in \cite{DGPS} for the SU(2) case. It is
given in a non-periodical gauge. To pass to the periodical gauge one
has to add $2\pi i\mu_m\delta_{mn}$ to $A_4$ (see the discussion at the end of the subsection D).
$A_4$ has the Coulomb-like form. In the periodical gauge
\beqa
A^{\rm per}_{4\,mn}&=&2\pi i \mu_m\delta_{mn}\delta_{\nu 4}
+\frac{i}{2}\delta_{mn}\left(\frac{1}{r_{m}}-\frac{1}{r_{m-1}}\right)\,,\\
A^{\rm per}_{i\,mn}&=&-\frac{i}{2}\delta_{mn}\left(\frac{1}{r_m}+\frac{1}{r_{m-1}}\right)
\sqrt{\frac{(\vrho_m-r_m+r_{m-1})(\vrho_m+r_m-r_{m-1})}
{(\vrho_m+r_m+r_{m-1})(r_m+r_{m-1}-\vrho_m)}}\;\;(e_{\varphi_m})_i
\eeqa
where $\vec e_{\varphi_m}\equiv\frac{\vec r_{m-1}\times \vec r_m}{|\vec r_{m-1}\times \vec r_m|}$.

\subsection{Reduction to the trivial holonomy case}

In the trivial holonomy limit ($\nu_L=1,\;\nu_{m\neq L}=0$) \eq{feq} simplifies.
It becomes a Shr\"odinger equation on the unit circle with only one delta function in
the left-hand-side. The solution is independent of $N$ and can be found in Ref.~\cite{KvBSUN}.
It is given by
\beq
f(\mu_m,\mu_n)\equiv f_0=\frac{\pi\sinh(2\pi r)}{\pi{\rho}^2\sinh(2\pi r)
+r\cosh(2\pi r)-r\cos(2\pi x_0)}
\eeq
where $2\pi\rho^2=\sum\rho_m$ and $r\equiv r_L$.

We now introduce a $N\times N$ unitary matrix $U$ which plays the role of the
`color orientation' of the (periodic) instanton to which the KvBLL reduces in the
trivial holonomy case. The first two columns of $U$ are defined through the spinors
\beq
U^m_n=\frac{1}{\rho}\zeta^m_\alpha,\quad n=\alpha=1,2.
\la{defU}\eeq
The rest columns are constrained only by the unitarity condition $U^\dagger U=1_N$;
they are not involved in the field construction. Correspondingly, the first two
rows of $U^\dagger$ are given by the hermitian conjugate spinors,
\beq
U^{\dagger\,n}_m=\frac{1}{\rho}\zeta^{\dagger\,\alpha}_m,\quad n=\alpha=1,2.
\la{defUdag}\eeq
This definition is non-contradictory if the two complex $N$-vectors $U^m_1$ and $U^m_2$ are
orthogonal and normalized to unity. Indeed, using \eqs{eq2}{size}, we obtain
\beq
\sum_{m=1}^N U^{\dagger\,\alpha}_m\,U^m_\alpha
=\frac{1}{\rho^2}\sum_{m=1}^N \zeta^{\dagger\,\alpha}_m\,\zeta^m_\alpha
=\frac{1}{2\pi\rho^2}\sum_{m=1}^N\left(\vrho_m\delta^\alpha_\beta
-\vec\vrho_m\cdot\vec\tau^\alpha_\beta\right)
=\delta^\alpha_\beta\,.
\la{unitarity}\eeq

To write down the gauge field of the trivial-holonomy instanton from the general
expressions \urs{Amu2}{defphi} we first replace there $\zeta\to U$ according to
\eqs{defU}{defUdag}:
\beqa
\la{ze1}
&&f_0\,\zeta^m_\alpha\zeta^{\dag\,\alpha}_n=f_0\rho^2\,(U\lambda^0 U^\dag)^m_n,\qquad
\d_\nu f_0\,\bar\eta_{\mu\nu}^a\,\zeta^m_\alpha(\tau^a)^\alpha_\beta\zeta^{\dag\,\beta}_n
=\d_\nu f_0\rho^2\,\bar\eta^a_{\mu\nu}\, (U\lambda^a U^\dag)^m_n,\\
\la{ze2}
&&\left(\phi^{-1}\right)^m_n=\delta^m_n-f_0\rho^2\, (U\lambda^0 U^\dag)^m_n
=U^m_\alpha(1-f_0\rho^2 \lambda^0)^\alpha_\beta U^{\dag\,\beta}_n,
\eeqa
where $(\lambda^0,\;\lambda^a)$ are $N\times N$ matrices with $(1_2,\;\tau^a)$ put
into the left-upper corner. The last term in \eq{Amu2} is again zero, and we arrive
at the compact result for the trivial-holonomy caloron:
\beq
\left(A_\mu\right)^m_n=\frac{1}{2}\bar\eta^a_{\mu\nu}\,(U\lambda^a U^\dag)^m_n\,\d_\nu\log\Pi
\la{Amu_small}\eeq
where
\beq
\Pi\equiv \frac{1}{1-f_0\rho^2}=1+\frac{\pi\rho^2\sinh(2\pi r)}{r[\cosh(2\pi r)-\cos(2\pi x_0)]}\,.
\eeq
This formula reproduces exactly the Harrington-Shepard instanton with arbitrary `gauge orientation'
$U$ defined, as we see from \eq{defU}, by the dyon relative coordinates $\vec\vrho_m$ and the relative
$U(1)$ orientation angles.

\subsection{Small-size KvBLL caloron}

Another important limit when the caloron field has a simple form is the case of small
$\rho=\sqrt{\sum_m\vrho_m/(2\pi T)}\ll 1/T$ implying that dyons' separations
are small, $\vrho_m\ll 1/T$ (in this subsection we restore the temperature
factors). We are interested in the caloron field at the distances $r$
from the center of the group of $N$ dyons, larger than the separations between them,
$r\sim \rho\gg \vrho_m$. Therefore, we can put in the leading order $r_m=r$ for all
$m=1,\dots,N$. We also consider the range of $x_4\sim\rho$ where the field is large.
In this range, the ADHM Green function is simply
\beq
f_{mn}=\frac{1}{r^2+x_4^2+\rho^2}\,.
\eeq
Repeating the calculations from the previous subsection we arrive at a standard expression
for the BPST instanton:
\beq
A_\mu=\frac{1}{2}\bar\eta^a_{\mu\nu}U\lambda^a U^\dag\d_\nu\log\Pi,\qquad
\Pi=1+\frac{\rho^2}{r^2+x_4^2},
\la{Amu_small_1}\eeq
where the instanton orientation matrix $U$ is given by \eq{defU}. Corrections
to \eq{Amu_small_1} die out as $T$ in the range $r\sim x_4\sim \rho\gg \vrho_m$ where
the field is large.

\Eq{Amu_small_1} is the approximate gauge field for small-size calorons in the non-periodic
gauge used in Ref.~\cite{KvBSUN}. To obtain the approximate small-$\rho$ field in the periodic gauge,
one has to gauge-transform \eq{Amu_small_1}:
\beq
{A^{\rm per}_\nu}_{mn}=2\pi i \mu_m\delta_{mn}\delta_{\nu 4}+(g^\dag A_\nu g)_{mn}
\eeq
where $g_{mn}=\delta_{mn}e^{2\pi i\mu_m x_4}$.

We note finally that when all dyons' separations $\vrho_m$ are small, the metric determinant
is given by \eq{M_triv} (even though the holonomy can be non-trivial!), and the caloron
measure coincides with that of the standard instanton, as shown in Section VII.

\section{Conclusions}

The metric of the $4N$-dimensional moduli space of the general $SU(N)$ caloron with
arbitrary eigenvalues of the Polyakov line at spatial infinity and at any temperature,
is given in terms of the spatial coordinates of the $N$ dyons that constitute the caloron,
and their $U(1)$ phases.

We have computed the determinant of the metric tensor, which defines the weight of the
$SU(N)$ caloron contribution to the partition function. The metric determinant
is a function of the $3d$ separations between dyons and of the Polyakov loop eigenvalues.
When all those eigenvalues are equal, it is the ``trivial holonomy'' case, and the KvBLL caloron
reduces to the usual periodic instanton whose moduli space is usually written in terms of the
instanton position, size and orientation. We have shown that the $SU(N)$ instanton measure
written in these variables coincides exactly with the one written in terms of the coordinates
and phases of the instanton constituents, the dyons. This result is not altogether
trivial, as in the first case the measure arises from the volume of the $SU(N)/SU(N-2)$
coset whereas in the second case it follows from the $3d$ geometry. We have also identified
the instanton $SU(N)$ orientation matrix through the dyons positions and $U(1)$ phases.

The following emerging physical picture may be plausible. The adequate degrees of freedom
in the Yang--Mills vacuum are, at any temperatures, calorons with non-trivial holonomy, which
are more general than the standard periodic instantons with trivial holonomy. The measure
should be described in terms of dyons' positions and phases. The free energy of the ensemble of
interacting dyons should be studied; hopefully at low temperatures it has a minimum at the
``most non-trivial holonomy" corresponding to $\Tr\,L=0$, however at $T>T_c$ related to $\Lambda$
there must be $N$ degenerate minima corresponding to trivial holonomy. An indication
that this may indeed be the case has been presented for $SU(2)$ in Ref.~\cite{DGPS}.
If correct, it would serve as the microscopic mechanism of the confinement-deconfinement
transition.

At low temperatures, although the correct description is still in terms of dyons with non-trivial
holonomy supporting the confinement, statistical fluctuations will lead to a large portion
of dyons that are not widely separated. If a group of $N$ different-type dyons happen to be
close to each other, the configuration is locally undistinguishable from the standard
$SU(N)$ instanton. Small-size instantons can be described both in the ``position--size--orientation"
terms, and in terms of dyons. However, for large-size overlapping instantons the former language
looses sense while the latter remains valid.

This physical picture (calling, of course, for a detailed mathematical study) may justify
the adequacy of the small-size instantons in physics related to the spontaneous chiral
symmetry breaking, while simultaneously explaining confinement as presumably due to dyons.\\

\noi
{\bf Acknowledgments}\\

We are grateful to Victor Petrov and Sergey Slizovskiy for helpful discussions.

\appendix

\section{Volume of the moduli space}

To get the moduli space volume, one has to integrate over the positions of $N$
dyons with the perimeter of the $N$-polygon fixed. More precisely we have to evaluate
the following integral
\beq
\mu_N(s)\equiv\int \prod_{m=1}^N d^3\vrho_m\det M\;\delta\!\left(\sum_{i=1}^N\vrho_m-s\right)
\;\delta^3\!\left(\sum_{m=1}^N\vec\vrho_m\right)
=\int \prod_{m=1}^{N}d^3\vrho_m\frac{4\pi^{N+1} s}{\prod_m\vrho_m}\;
\delta\!\left(\sum_{m=1}^N\vrho_m-s\right)\;\delta^3\!\left(\sum_{m=1}^N\vec\vrho_m\right)\,.
\la{defmu}\eeq
To reduce the number of integrations in eq.(\ref{defmu}) we use the following trick.
We introduce auxiliary integrals over Feynman parameters to reproduce the $\delta$-functions:
\beq
\mu_N(s)=
\int\prod_{m=1}^{N}d^3\vrho_m\,\frac{4\pi^{N+1} s}{\prod_m\vrho_m}\int\frac{d^3\alpha d\beta}{(2\pi)^4}
 \exp\left(i\left(\sum\vrho_m-s\right)(\beta+i\epsilon)+i\sum_{m=1}^N\vec\vrho_m\vec\alpha\right)\,.
\eeq
The infinitesimal $i\epsilon$ is added to ensure convergence.
Now we can integrate over $\vec\vrho_m$ since the integrals are factorized:
\beq
\mu_N(s)=
\int\frac{d^3\alpha d\beta}{(2\pi)^4}
\prod_{i=1}^{N}\left(\int\frac{d^3\vrho_m}{\vrho_m}e^{i\vrho_m(\beta+i\epsilon)+i\vec\vrho_m\vec\alpha}\right)
4\pi^{N+1} s\; e^{-i\beta s}\,.
\eeq
The $i\epsilon$ shift makes each integral over $\vrho_m$ finite. One can easily calculate it:
\beq
\int d\vrho_m d\cos\theta\;2\pi\vrho_m  e^{i\vrho_m(\beta+i\epsilon)}
e^{i\vrho_m\alpha \cos\theta}=\frac{4\pi}{\alpha^2-(\beta+i\epsilon)^2}\,.
\eeq
Now the measure can be written as a $4d$ integral
\beq
\mu_N(s)=
\int\frac{4\pi \alpha^2 d\alpha d\beta}{(2\pi)^4}
\frac{4(4\pi)^{N}\pi^{N+1}s }{[\alpha^2-(\beta+i\epsilon)^2]^N}
e^{-i\beta s}\,.
\label{muN}\eeq
From dimensions, $\mu_N(s)=\alpha_N s^{2N-3}$ where $\alpha_N$ is a constant to be computed;
we find it by induction. We first consider the $N=2$ case where $\mu_2(s)$ can be found
directly from \eq{defmu}:
\beq
\mu_2(s)=\int d^3\vrho_1\frac{8\pi^3}{\vrho_1}\delta(2\vrho_1-s)=2^3\pi^4 s\,.
\eeq
This implies $\alpha_2=2^3\pi^4$. For general $N$ we rotate the integration contour
$\alpha\rightarrow - i\alpha$ in eq.(\ref{muN}) since the poles are at $\pm(\beta+i\epsilon)$.
We can then rewrite \eq{muN} in an $SO(4)$ invariant form
\beq
\mu_N(s_\mu)=\frac{i s}{\pi}\int\frac{d^4\alpha}{(2\pi)^4}
\frac{(4\pi^2)^{N+1}(-1)^N}{(\alpha_\mu\alpha_\mu)^{N}}e^{-i\alpha_\mu s_\mu}
\eeq
where $\alpha_\mu=(\vec\alpha,\beta)$. The crucial step is the following
recurrent relation:
\beq
\frac{\mu_{N-1}(s_\mu)}{s}=\frac{1}{4\pi^2}\d^2_\mu\frac{\mu_{N}(s_\mu)}{s}
=\frac{\alpha_{N}}{4\pi^2}\d_\mu^2 s^{2N-3}
=\frac{\alpha_{N}}{4\pi^2}\frac{1}{s^3}\d_s \left(s^3\d_s s^{2N-4}\right)
=\frac{\alpha_{N}}{4\pi^2}(2N-4)(2N-2)s^{2N-6}
\eeq
where we have used that the radial part of the $4d$ Laplace
operator is $\d_\mu^2 f(s)=\frac{1}{s^3}\d_s \left(s^3 \d_s f(s)\right)$.
The solution to this equation is
\beq
\alpha_N=\frac{\pi^2\alpha_{N-1}}{(N-1)(N-2)}.
\eeq
Since $\alpha_2$ is known, it immediately follows that
\beq
\mu_N(s)=\frac{\alpha_2\pi^{2N-4}s^{2N-3}}{(N-1)!(N-2)!}=\frac{2^3\pi^{2N}s^{2N-3}}{(N-1)!(N-2)!}
\la{res1}
\eeq
which is used in section VI.

\section{Green function of the ADHM construction}

The Green function $f(z,z')$ is a very important object in the ADHM construction
and is used in many formulae. We derive here a compact expression for this key quantity.
An alternative expression for $f(z,z')$ can be found in Ref.~\cite{CKB}.
For the $SU(N)$ caloron it is defined by a Shr\"odinger equation on the unit circle~\cite{KvBSUN}:
\beq
\left[\left(\frac{1}{2\pi i}\d_z-x_0\right)^2+r(z)^2
+\frac{1}{2\pi}\sum_m\delta(z-\mu_m)\vro_m\right]f(z,z')=\delta(z-z')
\la{feq}\eeq
where $r(z)\equiv |\vec x-\vec y(z)|$.

To find $f(z,z')$ we first derive a closed system of linear algebraic equations
for $f_{mn}\equiv f(\mu_m,\mu_n)$. Assuming $f_{mn}$ and $f_{m+1,n}$ known we can
present $f(z,\mu_n)$ in the interval $(\mu_m,\mu_{m+1})$ in a standard way from solving \eq{feq}:
\beq
f(z,\mu_n)=-e^{2\pi i x_0 (z-\mu_m)}f_{mn}\frac{\sinh[2\pi r_m(z-\mu_{m+1})]}{\sinh(2\pi r_m\nu_m)}
+e^{2\pi i x_0 (z-\mu_{m+1})}f_{m+1,n}\frac{\sinh[2\pi r_m(z-\mu_{m})]}{\sinh(2\pi r_m\nu_m)}.
\eeq
Taking the derivatives near the discontinuity points one has
\beqa
\nn f'(\mu_m+\epsilon,\mu_n)&=&2\pi\left(i x_0 f_{mn}-r_m \coth(2\pi r_m\nu_m)f_{mn}
+e^{-2\pi i x_0\nu_m}\frac{r_m}{\sinh(2\pi r_m\nu_m)}f_{m+1, n}\right),\\
\nn f'(\mu_m-\epsilon,\mu_n)&=&2\pi\left(i x_0 f_{mn}+r_{m-1} \coth(2\pi r_{m-1}\nu_{m-1})f_{mn}
-e^{2\pi i x_0\nu_{m-1}}\frac{r_{m-1}}{\sinh(2\pi r_{m-1}\nu_{m-1})}f_{m-1,n}\right).
\eeqa
It follows from \eq{feq} that
\beq
-\frac{1}{4\pi^2}{\rm disc}f'(\mu_m,\mu_n)=\delta_{mn}-\frac{\vrho_m}{2\pi}f_{mn}
\eeq
and we can conclude that
\beq
f_{mn}={F^{-1}}_{mn}
\eeq
where
\beq
2\pi F_{mn}=\delta_{mn}\left[\coth(2\pi r_m\nu_m) r_m
+\coth(2\pi r_{m-1}\nu_{m-1}) r_{m-1}+\vrho_n\right]
-\frac{\delta_{m+1,n}r_m e^{-2\pi i x_0\nu_m}}{\sinh(2\pi r_m\nu_m)}
-\frac{\delta_{m,n+1} r_n  e^{2\pi i x_0 \nu_n}}{\sinh(2\pi r_n\nu_n)}\,.
\la{F}\eeq
Now we can reconstruct $f(z,z')$ for arbitrary $z$ and $z'$. We look for the solution
in the form
\beq
f(z,z')=s_m(z) f_{mn}s^{\dag}_n(z')+2\pi s(z,z')\delta_{[z][z']}
\la{fzz}\eeq
where $s(\mu_m,z')=0,\;s(\mu_n,z')=0$; we denot $[z]\equiv m$ if $\mu_m\leq z <\mu_{m+1}$.
The first term satisfies the homogeneous equation with given boundary conditions,
the second term gives $\delta(z-z')$ and vanishes at the boundary. The functions appearing
in \eq{fzz} are
\beqa
s_m(z)&=&e^{2\pi i x_0(z-\mu_m)}\frac{\sinh[2\pi r_m(\mu_{m+1}-z)]}{\sinh(2\pi r_m\nu_m)}\delta_{m[z]}
+e^{2\pi i x_0(z-\mu_{m})}\frac{\sinh[2\pi r_{m-1}(z-\mu_{m-1})]}
{\sinh(2\pi r_{m-1}\nu_{m-1})}\delta_{m,[z]+1},\\
s(z,z')&=&e^{2\pi i x_0(z-z')}\frac{\sinh\!\left(2\pi r_{[z]}
(\min\{z,z'\}-\mu_{[z]})\right)\sinh\!\left(2\pi r_{[z]}
(\mu_{[z]+1}-\max\{z,z'\})\right)}{r_{[z]}\sinh\!\left(2\pi r_{[z]}\nu_{[z]}\right)}.
\eeqa
\Eq{fzz} is convenient in some calculations since the main dependence on $z,z'$ is factorized.


\begin{thebibliography}{99}

\bibitem{BPST}
A. Belavin, A. Polyakov, A. Schwartz and Yu. Tyupkin, Phys. Lett. {\bf 59}, 85 (1975).

\bibitem{Pol}
A. Polyakov, Nucl. Phys. {\bf B120}, 429 (1977).

\bibitem{D96}
D. Diakonov, in: {\it Selected Topics in Non-Perturbative QCD}, eds.
A. Di Giacomo and D. Diakonov (Amsterdam, 1996) p. 397, {\tt hep-ph/9602375};\\
T. Sch\"{a}fer and E. Shuryak, Rev. Mod. Phys. {\bf 70}, 323 (1998), {\tt hep-ph/9610451}.

\bibitem{D02}
D. Diakonov, Prog. Part. Nucl. Phys. {\bf 51} (2003) 173, {\tt hep-ph/0212026}.

\bibitem{ILM}
C. Callan, R. Dashen and D. Gross, Phys. Rev. {\bf D17}, 2717 (1978);\\
E.M. Ilgenfritz and  M. M\"uller-Preussker, Nucl. Phys. {\bf B184}, 443 (1981);\\
E. Shuryak, Nucl. Phys. {\bf B203}, 93 (1982);\\
D. Diakonov and V. Petrov, Nucl. Phys. {\bf B245}, 259 (1984).

\bibitem{DP-SCSB}
D.~Diakonov and V.~Petrov, Phys. Lett. {\bf B147}, 351 (1984);
Nucl. Phys. {\bf B272}, 457 (1986);

\bibitem{Pol78}
A. Polyakov, Phys. Lett. {\bf 72}, 4770 (1978).

\bibitem{DP95}
D.~Diakonov and V.~Petrov, in: {\it Non-perturbative methods in Quantum
Chromodynamics}, ed. D.~Diakonov, PNPI, Gatchina (1995), p. 239;
D. Diakonov and V. Petrov, Phys. Scripta {\bf 61}, 536 (2000), {\tt hep-lat/9810037}.

\bibitem{HS}
B.J. Harrington and H.K. Shepard, Phys. Rev. {\bf D17}, 2122 (1978);
Phys. Rev. {\bf D18}, 2990 (1978).

\bibitem{GPY}
D.J.~Gross, R.D.~Pisarski and L.G.~Yaffe, Rev. Mod. Phys. {\bf 53}, 43 (1981).

\bibitem{DMir}
D.~Diakonov and A.~Mirlin, Phys. Lett. {\bf B203}, 299 (1988).

\bibitem{DP-Pol-line}
D.~Diakonov and V.~Petrov, unpublished (2000).

\bibitem{Zhitn}
S.~Jaimungal and A.~R.~Zhitnitsky,
{\tt hep-ph/9904377, 9905540}.

\bibitem{DMaul}
D. Diakonov and M. Maul, Nucl. Phys. {\bf B571}, 91 (2000), {\tt hep-th/9909078}.

\bibitem{KvB}
T.C.~Kraan and P.~van Baal, Phys. Lett. {\bf B428}, 268 (1998), {\tt hep-th/9802049};
Nucl. Phys. {\bf B533}, 627 (1998), {\tt hep-th/9805168}.

\bibitem{LL}
K.~Lee and C.~Lu, Phys. Rev. {\bf D58}, 025011 (1998), {\tt hep-th/9802108}.

\bibitem{KvBSUN}
T.~C.~Kraan and P.~van Baal,
Phys. Lett. {\bf B435}, 389 (1998), {\tt hep-th/9806034}.

\bibitem{BvB}
F.~Bruckmann, D.~Nogradi and P.~van Baal, Nucl. Phys. {\bf B698}, 233 (2004),
{\tt hep-th/0404210}.

\bibitem{BPS}
E.B. Bogomolnyi, Sov. J. Nucl. Phys {\bf 24}, 449 (1976) [Yad. Fiz. {\bf 24}, 861 (1976)]; \\
M.K. Prasad and C.M. Sommerfeld, Phys. Rev. Lett. {\bf 35}, 760 (1975).

\bibitem{DHKM}
N.M.~Davies, T.J.~Hollowood, V.V.~Khoze and M.P.~Mattis,
Nucl. Phys. {\bf B559}, 123 (1999), {\tt hep-th/9905015}; \\
N.M.~Davies, T.J.~Hollowood and V.V.~Khoze, {\tt hep-th/0006011}.

\bibitem{DP-SUSY}
D. Diakonov and V. Petrov, Phys. Rev. {\bf D67}, 105007 (2003), {\tt hep-th/0212018}.

\bibitem{DGPS}
D.~Diakonov, N.~Gromov, V.~Petrov and S.~Slizovskiy, Phys. Rev. {\bf D70}, 036003 (2004),
{\tt hep-th/0404042, hep-th/0407353}.

\bibitem{LWY}
K.M.~Lee, E.J.~Weinberg and P.~Yi,
Phys. Rev. {\bf D54}, 1633 (1996), {\tt hep-th/9602167}.

\bibitem{LY}
K. Lee and P. Yi, Phys. Rev. {\bf D56}, 3711 (1997), {\tt hep-th/9702107}.

\bibitem{Kraan}
T.C.~Kraan, Commun. Math. Phys. {\bf 212}, 503 (2000), {\tt hep-th/9811179};
Ph.D. Thesis, Leiden University (2000)
[available at http://www.lorentz.leidenuniv.nl/vanbaal/HOME/PUBL/kraan.ps].

\bibitem{ADHM}
M.F.~Atiyah, V.G.~Drinfeld, N.J.~Hitchin and Yu.I.~Manin, {\bf 65A}, 185 (1978);\\
N.H.~Christ, E.J.~Weinberg and N.K.~Stanton, Phys.~Rev. {\bf D18}, 2013 (1978);\\
E.~Corrigan, P.~Goddard and S.~Templeton, Nucl.~Phys. {\bf B151}, 93 (1979).

\bibitem{Nahm80}
W.~Nahm, Phys.~Lett. {\bf B90} (1980) 413.

\bibitem{Bernard}
C.~Bernard, Phys. Rev. {\bf D19}, 3013 (1979).

\bibitem{cal-lat}
R.C.~Brower, D.~Chen, J.~Negele, K.~Orginos and C.I.~Tan,
Nucl. Phys. Proc. Suppl.  {\bf 73}, 557 (1999), {\tt hep-lat/9810009};\\
E.M.~Ilgenfritz, B.V.~Martemyanov, M.~M\"uller-Preussker,
S.~Shcheredin and A.I.~Veselov,
Phys. Rev. {\bf D66}, 074503 (2002), {\tt hep-lat/0206004};
E.M.~Ilgenfritz, B.V.~Martemyanov, M.~M\"uller-Preussker and
A.I.~Veselov,
{\tt hep-lat/0402010};\\
C. Gattringer, Phys. Rev. {\bf D67}, 034507 (2003), {\tt hep-lat/0210001};
C. Gattringer and S. Schaefer, Nucl. Phys. {\bf B654}, 30 (2003), {\tt hep-lat/0212029};
C.~Gattringer {\it et al.}, Nucl.~Phys.~Proc.~Suppl. {\bf 129}, 653 (2004),
{\tt hep-lat/0309106};\\
F.~Bruckmann, E.M.~Ilgenfritz, B.V.~Martemyanov and P.~van~Baal,
Phys. Rev. {\bf D70}, 105013 (2004), {\tt hep-lat/0408004}.

\bibitem{CKB}
M.N.~Chernodub, T.C.~Kraan and P.~van Baal,
Nucl. Phys. Proc. Suppl. {\bf 83}, 556 (2000), {\tt hep-lat/9907001}.

\bibitem{GM}
G.W.~Gibbons and N.S.~Manton, Nucl. Phys. {\bf B274}, 183 (1986); Phys. Lett.
{\bf B356}, 32 (1995).

\bibitem{footnote1}
We add a common factor $8\pi^2$ as compared to Eq. (78) of~\cite{Kraan} such that
the normalization of the zero modes is in accordance with the previous subsection.

\end{thebibliography}
\end{document}